\documentclass{elsart}

\usepackage{epsfig}
\usepackage{amssymb,amsmath}
\usepackage{textcomp}
\usepackage{url}

\newcommand{\ecm}{\ensuremath{e~\text{cm}}}

\begin{document}

%\rightline{PSI-?? 07/??}

\begin{frontmatter}



\title{Compact storage ring to search for the muon electric dipole moment}


\author[PSI]{A.~Adelmann,} 
\author[PSI]{K.~Kirch,} 
\author[KVI]{G.J.G.~Onderwater,}
\author[PSI]{T.~Schietinger}
\address[PSI]{Paul Scherrer Institut, CH-5232 Villigen PSI, Switzerland}
\address[KVI]{Kernfysisch Versneller Instituut and University of Groningen, NL-9747AA Groningen, The Netherlands}

\vspace{1cm}

\begin{abstract}
We present the concept of a compact storage ring of less than 0.5~m orbit radius to search 
for the electric dipole moment of the muon ($d_\mu$) by adapting the ``frozen spin'' method.
At existing muon facilities a statistics limited sensitivity of 
$d_\mu \sim 5 \times 10^{-23}~\ecm$ can be achieved within one year of data taking.  
Reaching this precision would demonstrate the viability of this novel technique to
directly search for charged particle EDMs and already test a number of Standard Model extensions.  
At a future, high-power muon facility a statistical reach of $d_\mu \sim 5 \times 10^{-25}~\ecm$ 
seems realistic with this setup. 
\end{abstract}

\end{frontmatter}

% main text

%%%%%%%%%%%%%%%%%%%%%%%%%%%%%%%%%%%%%%%%%%%%%%%%%%%%%%%%%%%%%%%%%%%%%%%%%%%%%%%%%%%%%%%%%%%
\section{Motivation}
%%%%%%%%%%%%%%%%%%%%%%%%%%%%%%%%%%%%%%%%%%%%%%%%%%%%%%%%%%%%%%%%%%%%%%%%%%%%%%%%%%%%%%%%%%%

The observed matter-antimatter asymmetry of the Universe is not accounted for by the known 
extent of CP violation present in the Standard Model (SM) of particle physics.  
The search for permanent electric dipole moments (EDM) of fundamental particles is regarded 
as one of the most promising avenues for finding manifestations of additional CP violation, 
see, e.g.,~\cite{Pos05}.  
As these EDMs violate time reversal invariance (T) and parity (P), they also violate CP, if 
CPT invariance is assumed.  Various systems have been under investigation for a long time 
with limits becoming more and more restrictive, often challenging models beyond the SM.  
Of particular interest are the searches for the EDM of the neutron~\cite{Bak06}, the
electron~\cite{Reg02} and the $^{199}$Hg atom~\cite{Rom01}.

The muon is the only elementary particle for which the EDM ($d_\mu$) has been measured 
directly.  
Existing limits have been obtained parasitically at storage rings designed to measure the 
muon's anomalous magnetic dipole moment.  Currently the best limit is 
$1.8 \times 10^{-19}~\ecm$ at 95\% confidence level~\cite{Ben08}.  
This leaves the muon EDM as one of the least tested observables in the realm of the SM, 
which predicts a negligibly small value $<10^{-36}~\ecm$~\cite{Pos91}.  
For the muon there is no ongoing competitive dedicated search for its EDM.

Lepton-universality, together with the best current limit on the electron EDM, 
$d_e < 2.2 \times 10^{-27}~\ecm$~\cite{Reg02}, suggests a stringent limit on the muon EDM, 
$d_\mu \simeq (m_\mu/m_e) d_e < 5 \times 10^{-25}~\ecm$.  
There are, however, a number of models in which flavor-violating effects 
lead to a significant modification of this naive mass scaling (see, e.g., 
\cite{Bab00,Bab01,Fen01,Rom02,Pil02,Bab03,Bar03}). 
While the new limits on the $\tau \rightarrow \mu\gamma$ branching fraction 
set by the BABAR and Belle collaborations \cite{tauBF} call for a reappraisal 
of the predictions of these models, a muon EDM in the $10^{-22}~\ecm$ range 
still seems possible. 

Additional strong motivation to search for a muon EDM in the range $10^{-20}$ to 
$10^{-23}~\ecm$ arises from the result of the Brookhaven muon $(g-2)$ experiment, 
which challenges the SM prediction with a deviation of about three standard 
deviations~\cite{Ben06,Rob06}. 
First of all, it is well known (see, e.g.~\cite{Bai78}) and has been
re-emphasized~\cite{Fen01,Fen03} that the muon $(g-2)$ experiment by
itself cannot exclude a contribution of $d_\mu$ to the observed
precession frequency.
Amazingly, if the $(g-2)$ experiment observes a beyond-SM effect due
to new physics, parameterized as $a_\mu=a_\mu^\text{SM}+
a_\mu^\text{NP}$, with $a_\mu = (g_\mu-2)/2$, and $d_\mu \equiv
d_\mu^{\text{NP}}$, it could be entirely due to $d_\mu \sim
10^{-19}~\ecm$ and $a^\text{NP}=0$.  
Although a muon EDM as large as $10^{-19}~\ecm$ seems very unlikely, there is no solid 
theoretical argument against it either, and only an improvement of the experimental limit 
can settle the issue.

Secondly, it has been pointed out~\cite{Fen01} that if indeed $a_\mu^\text{NP} \neq 0$, 
as suggested by the Brookhaven measurement, one should generically also expect 
$d_\mu \propto a_\mu^\text{NP} \times \cos\phi_\text{CP}$, which for 
$\phi_\text{CP}\sim 1$ would result in $d_\mu \sim 10^{-22}~\ecm$, assuming the present 
values for $a_\mu$.  This situation calls for a dedicated experimental search for the 
muon EDM with a sensitivity of $10^{-22}~\ecm$ or better.

In this Letter we introduce an almost table top setup to perform such a search.  
It is based on a storage ring with an orbit of less than 0.5~m radius, and employs the 
so-called ``frozen spin'' technique introduced by Farley, et al.~\cite{Far04}.  
We show that this experiment would have an intrinsic sensitivity comparable to that of a
7~m radius ring proposed in the past~\cite{Sem00,Aok03}.  
We address injection into such a small ring and evaluate those systematic errors which 
depend on the muon momentum.  
We find that at existing muon beam facilities a sensitivity of $5\times 10^{-23}~\ecm$ 
can be reached in a year with ``one-muon-at-a-time''.  
This experiment would be statistics limited and could therefore be further improved by 
one or more orders of magnitude once new strong pulsed muon sources become available.

%%%%%%%%%%%%%%%%%%%%%%%%%%%%%%%%%%%%%%%%%%%%%%%%%%%%%%%%%%%%%%%%%%%%%%%%%%%%%%%%%%%%%%%%%%%
\section{Method and sensitivity}
%%%%%%%%%%%%%%%%%%%%%%%%%%%%%%%%%%%%%%%%%%%%%%%%%%%%%%%%%%%%%%%%%%%%%%%%%%%%%%%%%%%%%%%%%%%

The basic idea of the ``frozen spin'' method~\cite{Far04} is to cancel the regular ($g-2$) 
spin precession in a magnetic storage ring by the addition of a radial electric field.  
In the presence of a non-zero EDM, 
$d_\mu=\eta e \hbar/ 4mc\simeq \eta\times4.7\times10^{-14}~\ecm$, the spin will precess 
around the direction of the (motional) electric field,
\begin{equation}
  \label{Eq1}
  \vec{\omega}_e=\frac{\eta}{2}\frac{e}{m} \left( \vec{\beta}\times\vec{B} + \vec{E} \right)\,.
  \end{equation}
In the absence of longitudinal magnetic fields, the precession due to the anomalous 
magnetic moment is given by
\begin{equation}
  \label{Eq2}
  \vec{\omega}_a=\frac{e}{m}\left[a\vec{B}-\left(a-\frac{1}{\gamma^2-1}\right)\vec{\beta} \times \vec{E}\right] 
\end{equation}
with $a=(g-2)/2$.  One can reduce $\omega_a$ to zero by choosing
\begin{equation}
  \label{Eq3}
  E = \frac{aB\beta}{1-(1+a)\beta^2} \simeq aB\beta\gamma^2\,,
\end{equation}
for $a\beta^2\gamma^2\ll 1$.  Then, the only precession is the one of Eq.~(\ref{Eq1}) and 
the spin is ``frozen'' in case $d_\mu\!=\!\eta\!=\!0$.  
For $d_\mu \neq 0$, the muon spin, initially parallel to the muon momentum, is moving steadily 
out of the plane of the orbit.

The observable in the experiment is the up--down counting asymmetry $A$ due to the muon 
decay asymmetry.  
In the following discussion ``positron'' will refer to both electrons and positrons originating
from muon decays.  With polarization $P$, lifetime $\tau$ and the number of detected decay 
positrons $N$, the uncertainty in $\eta$ is to good approximation given by
\begin{equation}
  \label{Eq4}
  \sigma_\eta = \frac{\sqrt{2}}{\gamma \tau (e/m) \beta B A P \sqrt{N}}\,,
\end{equation}
which suggests that to obtain the best accuracy it is desirable to use a high magnetic field and 
high energy muons.  But according to Eq.~(\ref{Eq3}) this would require impractically large 
electric fields.  
Expressing Eq.~(\ref{Eq4}) in terms of $E$ from Eq.~(\ref{Eq3}),
\begin{equation}
  \label{Eq4new}
  \sigma_\eta = \frac{\sqrt{2}a c \gamma}{\tau (e/m) E A P \sqrt{N}}\,,
\end{equation}
we see that the boundary condition of a practically limited electric field strength actually 
favors {\em low} values of $\gamma$. 

Consequently, we consider the use of a low-momentum muon beam and, as a concrete example, 
the PSI \textmu E1 beamline~\cite{PSIbeamwebsite}. 
Depending on the mode of operation, one obtains up to 
$2\times 10^{8}$~s$^{-1}$ $\mu^+$ with $p_\mu\!=\!125$~MeV/$c$
($\beta\!=\!0.77$, $\gamma\!=\!1.57$) from backward decaying pions with $p_\pi\!=\!220$~MeV/$c$.  
The muons arrive in bunches every 19.75~ns (corresponding to the accelerator frequency) with a 
burst width slightly below 4~ns \cite{Bar00}.  
The muon polarization is $P\!\simeq\!0.9$, for the decay asymmetry we use $A\!=\!0.3$. 

We consider a scenario with magnetic and electric fields of $B=1$~T and $E=0.64$~MV/m, respectively, 
corresponding to a ring radius of $R=0.42$~m.  
Choosing a moderate value for the $B$-field allows the use of a normal conducting magnet to 
switch the field polarity reasonably fast.  
The muon momentum and the strength of the magnetic field fix the electric field strength at 
a value that can be readily achieved.  
Our choice of parameters results in an intrinsic sensitivity of
\begin{equation}
  \label{Eq4neweval}
  \sigma_{d_\mu} \simeq 1.1\times 10^{-16}~\ecm/\sqrt{N}\,.
\end{equation}
This is comparable to that presented in Ref.~\cite{Far04},
$\sigma_{d_\mu} \simeq 2\times 10^{-16}~\ecm/\sqrt{N}$ based on
$p_\mu=0.5$~GeV/$c$, $E=2.2$~MV/m, $B=0.25$~T, $\gamma=5$,
$R=7$~m, $P=0.5$.

The idea for the operation of the experiment at the  \textmu E1 beam line is to use one 
muon at a time in the storage ring and observe its decay before the next muon is injected.  
This way, the high beam intensity is traded off for beam quality and muons suitable for 
the injection can be selected.  
Assuming an injection latency of 1~\textmu s and an average observation time of 
$\gamma\,\tau_\mu=3.4$~\textmu s results in more than $2\times10^5$ muon decays per second 
and allows for $N \sim 4\times10^{12}$ detected events per year, thus
\begin{equation}
  \sigma_{d_\mu} \simeq 5\times 10^{-23}~\ecm\,.
\end{equation}

%%%%%%%%%%%%%%%%%%%%%%%%%%%%%%%%%%%%%%%%%%%%%%%%%%%%%%%%%%%%%%%%%%%%%%%%%%%%%%%%%%%%%%%%%%%
\section{Storage ring injection}
%%%%%%%%%%%%%%%%%%%%%%%%%%%%%%%%%%%%%%%%%%%%%%%%%%%%%%%%%%%%%%%%%%%%%%%%%%%%%%%%%%%%%%%%%%%

Injection into a compact storage ring is a significant challenge.  
In our application, the velocity of the 125~MeV/$c$ muons is about 23~cm/ns, corresponding 
to a revolution time of about 11~ns.  
The use of a conventional kicker device faster than the revolution time may not be feasible.  
Existing devices are at least an order of magnitude slower, although there are promising 
developments for the International Linear Collider (ILC)~\cite{Nai07}.  
An alternative and viable scheme for particle injection would be through beam resonance. 
Injection of electrons into small storage rings using 1/2 (and also 2/3) integer resonances 
has been demonstrated~\cite{Yam03,Has03} and can be adapted for muons.

This injection method is a time reversal of half integer resonance extraction~\cite{Tak87}.  
In the radial phase space the separatrix of the half integer resonance together with a 
stable region around the central orbit is created by a so-called perturbator.  
The perturbator is creating odd multipole fields with field strengths depending on the
radial betatron frequency.  
The muons are injected near the separatrix in the unstable region through an inflector.  
By ramping down the perturbator field, these muons are captured by expanding the stable
region of the phase space.

\begin{figure}[ht]
  \includegraphics[angle=0,scale=0.40]{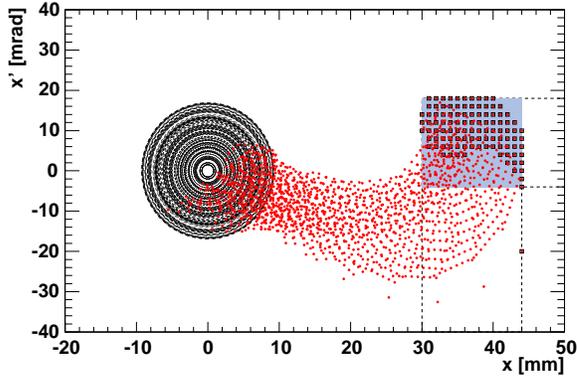}
  \caption{Simulation of twenty-turn muon resonance injection (red) from a
           collimated phase space (blue) ending in the stable region
           of the phase space.}
\label{fig:muinj1}
\end{figure}

The \textmu E1 beam line at PSI delivers muons in a transverse phase space of approximately 
$28 \times 17$~mm$\cdot$mrad when operating within 1\% of momentum acceptance (FWHM).  
The phase space can be reduced by a suitable collimation system to fit the acceptance of the
resonance injection scheme.  
We consider a weak focusing storage ring with a tune $\nu_x \sim 1/2$ .  
In Fig.~\ref{fig:muinj1} we show  the results of a simulation demonstrating twenty-turn 
injection (red) out of a narrow, i.e., collimated phase space (blue). 
The stable part of the phase space, i.e., the observation phase of the muons, is shown
in black.

The synchronization of the field ramping and the muon injection can be achieved via 
triggering on an upstream muon entrance telescope, in combination with the accelerator 
radio-frequency and the detection of the previous decay positron (or a suitable time-out).  
The loss in statistics due to the time needed for ramping up the perturbator is of order 10\%.  
The decrease of the actual observation time window due to the time needed for reaching the 
stable orbit is of the same order.   
Thus a total loss of statistics of about 15\% is expected.

The necessity to synchronize the perturbator ramping with a ``good'' muon comes from the 
low intensity at the existing muon beam (here the chance to have an acceptable muon in a 
20~ns time window is only about 2\%).  
Assuming a pulsed muon source of much larger intensity, the ramping of the perturbator can 
be synchronized to the machine frequency, which simplifies the injection.  
Many (e.g., $10^4$) muons within one bunch would then be captured into the stable orbit.

%%%%%%%%%%%%%%%%%%%%%%%%%%%%%%%%%%%%%%%%%%%%%%%%%%%%%%%%%%%%%%%%%%%%%%%%%%%%%%%%%%%%%%%%%%%
\section{Polarimetry}
%%%%%%%%%%%%%%%%%%%%%%%%%%%%%%%%%%%%%%%%%%%%%%%%%%%%%%%%%%%%%%%%%%%%%%%%%%%%%%%%%%%%%%%%%%%

The muon spin orientation is reconstructed from the distribution of the decay positrons.  
Due to the magnetic field of the storage ring, the positrons will be bent towards the inner 
side of the ring.   
Both the efficiency for detecting a decay positron and the analyzing power, i.e., 
sensitivity to muon polarization, must be optimized.

The simplest and most straightforward detection system only distinguishes upward and 
downward going positrons.  The number of upward versus downward going positrons is 
independent of the muon energy.  
In this case, the analyzing power for a vertical muon spin component is typically 
$A\simeq 0.3$.  
Efficiencies can be several tens of percent.

It was shown in \cite{Ben08} that detecting the vertical positron angle is less prone to 
systematic errors.  
The additional information on the positron also improves the statistical sensitivity of the
experiment.  
Contrary to the up--down counting asymmetry, the vertical angle does depend on the muon 
energy and is inversely proportional to $\gamma_\mu$.  Because the width of the vertical 
angle distribution has the same dependence, the relative precision does not depend on the
muon momentum.  
As a guard against systematic errors, however, a larger signal and thus a lower muon 
momentum is preferred.

%%%%%%%%%%%%%%%%%%%%%%%%%%%%%%%%%%%%%%%%%%%%%%%%%%%%%%%%%%%%%%%%%%%%%%%%%%%%%%%%%%%%%%%%%%%
\section{Systematic effects and countermeasures}
%%%%%%%%%%%%%%%%%%%%%%%%%%%%%%%%%%%%%%%%%%%%%%%%%%%%%%%%%%%%%%%%%%%%%%%%%%%%%%%%%%%%%%%%%%%

Two categories of systematic errors can be distinguished:  (1) those that lead to an 
{\em actual} growth of the polarization into the vertical plane; and (2) those that lead 
to an {\em apparent} vertical polarization component.  
In \cite{Far04}, the six dominant effects and their counter measures are discussed.  
The setup described in this Letter does not introduce additional systematic error sources, 
so that these counter measures are applicable to our setup as well.  
We briefly discuss those systematic errors which are affected by the lower muon momentum.

An important source of systematic error is the existence of an average electric field 
component $E_v$ along the magnetic field.  
The resulting false EDM is
\begin{equation}
  \eta_{\text{false}} \simeq \cfrac{2a^2\gamma^2}{\beta}\,\cfrac{E_v}{E_r}.
\end{equation}
At our momentum, the increased sensitivity due to a lower $\beta$ is more than compensated 
by the low $\gamma$:  $\gamma^2/\beta = 3.1$, as compared to the experiment suggested in, 
e.g., \cite{Aok03}, for which $\gamma^2/\beta = 23.9$.  
A systematic error at the level of the statistical reach ($\eta = 1\times 10^{-9}$) leads 
to the (modest) requirement $E_v/E_r < 1\times 10^{-4}$.  
Furthermore, when switching from clockwise to counter-clockwise injection, the false EDM 
remains the same, whereas the true EDM signal changes sign~\cite{Far04}.

A net longitudinal magnetic field $B_L = \frac{1}{2\pi\rho}\oint \vec{B}\cdot \vec{dl}$, 
combined with an initial transverse polarization $P_T$ leads to a false EDM of order
\begin{equation}
  \eta_{\text{false}} \simeq \cfrac{2\gamma}{\beta}\,\cfrac{B_L}{B_T}\,\cfrac{P_T}{P_L}\, .
\end{equation}
Assuming an initial transverse-to-longitudinal polarization $P_T/P_L$ of 10\% leads to the 
requirement $B_L/B_T < 6\times 10^{-9}$ to match the statistical uncertainty.  
The latter corresponds to a current of 13~mA flowing through the orbit of the stored muons, 
or an electric field change of 2.5~GV/m/s perpendicular to the orbit and synchronized to 
the measuring cycle.  Variation of the initial transverse polarization, or allowing a slow 
residual precession will expose this component. 

Since the experimental setup is quite small, shimming undesired field component to an 
acceptable level will be considerably easier than in a larger setup.  
Moreover, to match the relatively modest statistical reach leads to rather relaxed 
requirements on these field perturbations.

Systematic errors of the second category include shifts and rotations of the detectors.  
The optimal measuring cycle is about two lifetimes and thus scales with $\gamma$.  
For $\eta = 1\times 10^{-9}$, the growth of the up--down counting asymmetry $\alpha$ is 
$d\alpha/dt \simeq 1\times 10^{-6}/\gamma\tau$.  
Static detector and beam displacements therefore cannot lead to a false EDM signal.  
The effect of random motion, i.e., not correlated with the measurement cycle, is reduced 
by six orders of magnitude (the square-root of the number of measuring cycles), and 
therefore also not of any concern.  Only if the motion is synchronous with the measuring 
cycle, false EDM signals may appear.

The detector position relative to the average muon decay vertex determines the systematic 
error, so both detector and beam motion must be considered.  
Especially the latter is of concern, because it is intrinsically synchronized with the 
measurement.  
For a displacement $\delta$ along $\vec{B}$, the resulting false EDM when using the 
up--down counting ratio is
\begin{equation}
  \eta_{\text{false}} \simeq \frac{\gamma\tau}{10^3}\,\frac{d}{dt}\left( \frac{\delta}{l} \right)\,,
\end{equation}
with $l = \mathcal{O}(10~\text{cm})$ the typical scale of the experimental setup.  
This places the rather stringent limit $d\delta/dt <$\,0.1\,\textmu m/$\gamma\tau$.  
The positron momentum dependences of the true and this false EDM signal are significantly 
different so that they can be disentangled.

%%%%%%%%%%%%%%%%%%%%%%%%%%%%%%%%%%%%%%%%%%%%%%%%%%%%%%%%%%%%%%%%%%%%%%%%%%%%%%%%%%%%%%%%%%%
\section{Conclusion}
%%%%%%%%%%%%%%%%%%%%%%%%%%%%%%%%%%%%%%%%%%%%%%%%%%%%%%%%%%%%%%%%%%%%%%%%%%%%%%%%%%%%%%%%%%%

We have described a compact muon storage ring based on a novel resonant injection scheme as 
a viable setup to measure the EDM of the muon using the frozen spin technique.  
Such a measurement would demonstrate the feasibility of a still unexplored technique for the
direct search for EDMs of charged particles and would serve as a stepping stone for future 
applications of this promising method.

At existing muon facilities (PSI \textmu E1 beamline~\cite{PSIbeamwebsite}) a sensitivity 
of $d_\mu \sim 5 \times 10^{-23}~\ecm$ seems reachable in one year of data taking, an
improvement of the existing limit by more than three orders of magnitude.  
Already at this level of precision, several interesting physics tests are possible.  
First, it could \emph{unambiguously} exclude the EDM to be the explanation of the difference 
between the measured anomalous magnetic moment and its SM prediction.  
It would furthermore test various SM extensions, in particular those that do not respect 
lepton universality. 

In view of the possible advent of new, more powerful pulsed muon sources, the same 
experimental scheme can be realized but with considerably more muons per bunch being 
injected into the ring.  
It appears realistic to expect accelerators with on the order of 100~kHz repetition rates 
and more than $10^4$ muons stored per bunch.  
The statistical sensitivity of the described approach would then reach down to 
$d_\mu \sim 5 \times 10^{-25}~\ecm$ or better.  
Although systematic issues at this level of precision have been discussed in some detail 
in~\cite{Far04}, more detailed studies would be needed.

%%%%%%%%%%%%%%%%%%%%%%%%%%%%%%%%%%%%%%%%%%%%%%%%%%%%%%%%%%%%%%%%%%%%%%%%%%%%%%%%%%%%%%%%%%%
\section*{Acknowledgements}

We are grateful to 
M.~B\"oge,
W.~Fetscher,
K.~Jungmann,
S.~Ritt,
and A.~Streun
for fruitful discussions. 
Furthermore, we acknowledge that J.P.~Miller independently suggested the use of low-momentum 
muons to exploit the ``frozen-spin'' method.
The work by C.J.G.O. is funded through an Innovational Research Grant 
of the Netherlands Organization for Scientific Research (NWO).

%%%%%%%%%%%%%%%%%%%%%%%%%%%%%%%%%%%%%%%%%%%%%%%%%%%%%%%%%%%%%%%%%%%%%%%%%%%%%%%%%%%%%%%%%%%

\end{document}